\documentclass[11pt]{article}

\usepackage{overcite}
\usepackage{graphics}
\usepackage{epsfig}
\usepackage{amsmath}

\begin{document}

{\LARGE 
\begin{center}
{Shape oscillation of a levitated drop in an acoustic field}
\end{center}}

\begin{center}
W. Ran, \ S. Fredericks, \& J.R. Saylor\\
Clemson University \\
Department of Mechanical Engineering \\
Clemson, SC 29634 \\
\end{center}

\begin{abstract}
A `star drop' refers to the patterns created when a drop, flattened by some force, is excited into shape mode oscillations. These patterns are perhaps best understood as the two dimensional analog to the more common three dimensional shape mode oscillations. In this fluid dynamics video an ultrasonic standing wave was used to levitate a liquid drop. The drop was then flattened into a disk by increasing the field strength. This flattened drop was then excited to create star drop patterns by exciting the drop at its resonance frequency. Different oscillatory modes were induced by varying the drop radius, fluid properties, and frequency at which the field strength was modulated.
\end{abstract}

\section{Background}
\label{section:background}

When a drop is held within an acoustic field, there is a force balance between the pressure exerted by the field, which tends to flatten the drop, and the surface tension of the drop, which tends to form the drop into a sphere. Under stationary conditions this will result in an oblate shaped drop residing within the field. However, if the field strength is periodically modulated, an instability in the drop can occur. This will result in the formation of a wave which extends radially around the drop, orthogonal to the direction of the applied field. Visually, the result appears as a star shape, with several elliptical lobes extending from the center of the drop.

The number of peaks in the produced wave can be controlled by exciting the drop at various harmonics of the drop's resonance frequency.

\begin{equation}
\centering
f_n = \frac{1}{2\pi}\sqrt{\frac{n(n-1)(n+2)\gamma}{\rho R^3}},
\label{eq:lambFrequency}
\end{equation}
where $\rho$ is the liquid density, $\gamma$ is the surface tension, $R$ is the drop radius, and $n$ is the harmonic of the oscillation\cite{lamb1932}.

Various methods can be used to oscillate a liquid drop and form these star drops\cite{brunet2011}. These methods include vibration of a non-wetting substrate, pulsating airflow, and modulating the field strength of a supporting magnetic or acoustic field.


\section{Method}
\label{section:exp}

To produce the presented visualization, an ultrasonic standing wave field was created between an ultrasonic transducer and a reflector. At the nodes of this acoustic field it is possible to insert a liquid drop and levitate it.\cite{trfinh1985} The strength of the acoustic field is highly sensitive to the amplitude and frequency of the ultrasonic wave. To provide an initial perturbation to the drop the strength of the acoustic field was modulated through the use of frequency modulation. By matching the resonance frequency, $f_n$, the presented star patterns were formed\cite{shen2010}.


\section*{Acknowledgments}
\label{section:acknowledgements}

We would like to thank NIOSH, NSF and Dr. R. Glynn Holt of Boston University, Department of Mechanical Engineering for their support of this project.



\begin{thebibliography}{40}
\expandafter\ifx\csname natexlab\endcsname\relax\def\natexlab#1{#1}\fi
\providecommand{\bibinfo}[2]{#2}
\ifx\xfnm\relax \def\xfnm[#1]{\unskip,\space#1}\fi

\bibitem{lamb1932}
\bibinfo{author}{Lamb, H.}
	(\bibinfo{year}{1932}).
	\newblock \bibinfo{title}{{Hydrodynamics (6th Ed.)}}.
	\newblock {\it \bibinfo{journal}{Cambridge Univ. Press, Cambridge {UK}}\/}.

\bibitem{brunet2011}
\bibinfo{author}{Brunet, P.}, \bibinfo{author}{Snoeijer, J.H.}
  (\bibinfo{year}{2011}).
	\newblock \bibinfo{title}{{Star-drops formed by periodic excitation and on an air cushion - A short review}}.
	\newblock {\it \bibinfo{journal}{The European Physical Journal Special 	Topics}\/},  {\it
  \bibinfo{volume}{192.1}\/}, \bibinfo{pages}{207-226}.
	
	
	


\bibitem{trfinh1985}
\bibinfo{author}{Trinh, E.H.}
	(\bibinfo{year}{1985}).
	\newblock \bibinfo{title}{{Compact acoustic levitation device for studies in fluid dynamics and material science in the laboratory and microgravity}}.
	\newblock {\it \bibinfo{journal}{Review of Scientific Instruments}\/}, 		{\it \bibinfo{volume}{56}\/}, \bibinfo{pages}{2059-2065}.	


\bibitem{shen2010}
\bibinfo{author}{Shen, C.L.}, \bibinfo{author}{Xie, W.J.}, \bibinfo{author}{Wei, B.}
  (\bibinfo{year}{2010}).
	\newblock \bibinfo{title}{{Parametrically excited sectorial oscillation of liquid drops floating in ultrasound}}.
	\newblock {\it \bibinfo{journal}{Physical Review E}\/},  {\it
  \bibinfo{volume}{81.4}\/}, \bibinfo{pages}{046305. 6 pages}.

\end{thebibliography}
\end{document}